\newcommand{\FeH}{\hbox{\rm [Fe/H]}}	   
\newcommand{\aH}{\hbox{\rm [$\alpha$/Fe]}} 

\newcommand{\eg}{e.g.}			
\newcommand{\ie}{i.e.}			
\newcommand{\etal}{et~al.}		

\newcommand{\sect}{Sec.}		

\documentstyle[11pt,aaspp4]{article}
\tighten
  
\received{}
\revised{}
\accepted{}
\journalid{}{}
\articleid{}{}
\paperid{}

\lefthead{Holland}
\righthead{M31 Distance}

\slugcomment{to appear in the {\sl AJ}}

\begin{document}

\title{The Distance to the M31 Globular Cluster System}

\author{Stephen Holland}

\affil{Institut for Fysik og Astronomi \\
Aarhus Universitet \\
Ny Munkegade, Bygning 520 \\
DK--8000 {\AA}rhus C \\
Denmark}

\begin{abstract}
	The distance to the centroid of the M31 globular cluster
system is determined by fitting theoretical isochrones to the observed
red-giant branches of fourteen globular clusters in M31.  The mean
true distance modulus of the M31 globular clusters is found to be
$\mu_0 = 24.47 \pm 0.07$ mag.  This is consistent with distance moduli
for M31 that have been obtained using other distance indicators.
\end{abstract}



\section{Introduction\label{SECTION:intro}}

	Over the last ten years the distance to the Andromeda Galaxy
($=$~M31~$=$~NGC~224) has been measured using a variety of distance
indicators.  Pritchet~\&~van~den~Bergh\markcite{PV87}~(1987) used
RR~Lyrae variables to derive a true distance modulus of $24.34
\pm 0.15$.  The observed brightnesses of red giants in the halo of M31
suggest that $\mu_0 = 24.23 \pm 0.15$
(Pritchet~\&~van~den~Bergh\markcite{PV88}~1988).  Novae light curves
give a distance modulus of $24.27 \pm 0.20$
(Capaccioli~\etal\markcite{CD89}~1989) while Cepheid variables suggest
that $\mu_0 = 24.43 \pm 0.06$
(Freedman~\&~Madore\markcite{FM90}~1990).
Brewer~\etal\markcite{BR95}~(1995) found $\mu_0 = 24.36 \pm 0.03$
using carbon stars in the disk of M31 yet
Ostriker~\&~Gnedin\markcite{OG97}~(1997) found $\mu_0 = 24.03 \pm
0.23$ from the peak of M31's globular cluster (GC) luminosity
function.  A weighted mean of these six values yields
$\langle\mu_0\rangle = 24.36
\pm 0.03$ mag where the uncertainty is the standard error in the mean.
This corresponds to a distance of $\sim 745$ kpc.

	Recently Feast~\&~Catchpole\markcite{FC97}~(1997) used {\sl
Hipparcos\/} data to calibrate the Cepheid period--luminosity (PL)
relation and found a distance modulus of $\mu_0 = 24.77 \pm 0.11$ for
M31.  This corresponds to a distance of $\sim 900$ kpc, an increase of
$\sim 20$\% over the previously accepted value.  However,
Madore~\&~Freedman\markcite{MF97}~(1997) analysed the {\sl
Hipparcos\/} data and found that the pre-{\sl Hipparcos\/} zero point
for the Cepheid PL relation is good to within $\pm 0.14$ mag ($\sim
7$\%).  They also note that uncertainties due to reddening,
metallicity effects on the PL relation, and statistical uncertainties
in the data make up a significant component of the total uncertainty
in the PL zero point.

	This study determines an estimate of the distance to M31 by
finding the distance to the centroid of the M31 GC system.  The shape
of the red-giant branch (RGB) in a GC is strongly sensitive to the
metallicity of the GC (see Da~Costa~\&~Armandroff\markcite{DC90}~1990
for a discussion of this effect in Galactic GCs).  If the metallicity
of the GC is determined in a manner that is {\sl independent of the
distance to the GC}, and the amount of reddening along the
line-of-sight to the GC is known, then a reference isochrone of the
correct metallicity can be shifted in distance until the best fit to
the GC's RGB is obtained.


\section{The Data\label{SECTION:data}}

\subsection{The Red-Giant Branches\label{SECTION:rgb}}

	The RGB data used in this study were taken from three sources.
RGB fiducial sequences for G1, G58, G105, G108, G280, G219, and G351
were taken from Table 2 of Fusi~Pecci~\etal\markcite{FP96}~(1996,
hereafter referred to as FP96).  These fiducial sequences represent the
observed mean points of the RGBs {\sl before\/} reddening and
extinction corrections were applied.  The fiducial sequences for G302
and G312 were obtained by drawing mean lines through the RGBs of each
GC using the data from Holland~\etal\markcite{HF97}~(1997, hereafter
referred to as HFR).  The data for these two GCs were recalibrated
using the procedure described in HFR but with no reddening or
extinction corrections applied.  The fiducial sequences for G11, G319,
G323, G327, and G352 were obtained by drawing mean lines through the
RGBs of each GC using the data in Couture~\etal\markcite{CR95}~(1995,
hereafter referred to as CR95).  The fiducial sequences for these five
GCs were dereddened, and extinction corrections removed, by inverting
the reddening and extinction corrections described in CR95.

	FP96 used $B$-, $V$-, and $I$-band photometry from the
Wide-Field Planetary-Camera 2 (WFPC2) aboard the {\sl Hubble Space
Telescope\/} ({\sl HST\/}) to construct their color--magnitude
diagrams (CMDs).  HFR used WFPC2 $V$- and $I$- band photometry while
CR95 used $V$- and $I$-band photometry from the High Resolution Camera
at the Canada--France--Hawaii telescope (CFHT).  The data for G280 and
G351 are from $B$- and $V$-band photometry with the Faint Object
Camera aboard the {\sl HST}.  The limiting magnitudes (the $V$-band
magnitude where the published uncertainties in the photometry for the
individual stars is $\sigma_V \sim 0.1$) for the {\sl HST\/} CMDs are
$V_{\lim} \sim 26$.  For the CFHT data $V_{\lim} \sim 23$.  The
fiducial sequence for Bo468 (from FP90) was not used since there is no
published estimate of its iron abundance that was obtained using a
method that does not require prior knowledge of the distance to the
GC\@.

	Uncertainties were assigned to the fiducial points based on
the width of the observed RGB at each fiducial point.  The width of
the RGB was measured and assumed to be 1.5 times the Full-Width at
Half-Maximum (\ie\ the bin has a half-width of $\sim 3.5\sigma$) of
the scatter around the mean point.  This assumes that the scatter is
due to Gaussian uncertainties in the photometry.  It ignores
complications due to confusion between RGB and asymptotic giant branch
stars, binary stars, and contamination from stars in the halo of M31.
This means that the uncertainties can not be used as a measure of the
quality of individual fiducials.  However, they can be used to provide
an estimate of the relative weight to give each fiducial point {\sl in
a single GC's RGB}.

\subsection{The Iron Abundances\label{SECTION:FeH}}

	The iron abundances used in this study were taken from the
Huchra~\etal\markcite{HB91}~(1991) catalogue of iron abundances and
radial velocities for 150 GCs in the M31 system.  These iron
abundances were determined using six absorption lines indices measured
from integrated spectra of each GC and a calibration that was tied to
the metallicity scale of Zinn~\&~West\markcite{ZW84}~(1984).  The
method is fully described in Brodie~\&~Huchra\markcite{HB90}~(1990).
These iron abundances are completely independent of the distance to
M31, the amount of interstellar reddening, and the stellar photometry
used to generate the RGB fiducial sequences for the GCs.  Other
methods of estimating the iron abundances of M31 GCs, such as
integrated colors ({\eg}~B{\`o}noli~\etal\markcite{BF87}~1987) and
CMD-based methods ({\eg}~HRF), require a knowledge of either the
amount of reddening along the line-of-sight to the GC, or the distance
to M31, or both.  Therefore only spectroscopic estimates of the iron
abundance are used in this study.  FP96 found that
Huchra~\etal's\markcite{HB91}~(1991) spectroscopic iron abundances
were consistent with the iron abundances obtained from integrated
colors for all of the GCs in their sample except G280, assuming a
distance modulus of $\mu_0 = 24.43$ for M31.

\subsection{The Isochrones\label{SECTION:isochrones}}

	In order to determine the distance to each GC the fiducial RGB
sequences for each GC were compared to theoretical isochrones.  This
procedure is described in greater detail in
\sect~\ref{SECTION:results}.  The oxygen enhanced isochrones of
VandenBerg\markcite{V97}~(1997), with an age of $t_0 = 14$ Gyr, were
used.  This assumes that the M31 GC system has the same age as the
Galactic GC system.  The choice of age is not critical since location
of the RGB shifts by only $d(V\!-\!I)/dt_0 \sim 0.008$ mag/Gyr at $M_V
\sim -1$, which is small compared to the uncertainties in the
reddening and in the spectroscopic iron abundance for each GC\@.  An
alpha-element enhancement of $\aH = +0.3$ was adopted since this
amount of alpha-element enhancement is seen in Galactic GCs with a
wide range of iron abundances (see Carney\markcite{C96}~1996).  The
color transformations used to transform the isochrones from the
$(M_{\rm bol},T_{\rm eff})$ plane to the $(M_V,V\!-\!I)$ plane give
good fits up to $\FeH \sim -0.8$ but may be less reliable for higher
iron abundances (VandenBerg, private communication).

	For each GC an isochrone of the appropriate iron abundance was
obtained by interpolating across a grid of reference isochrones.  This
is not a formally correct method of obtaining isochrones with specific
iron abundances.  However, none of the interpolations were more than
0.08 dex away from a reference isochrone so the error introduced by
the interpolation will be small compared to the uncertainties in the
spectroscopic iron abundance measurements ($\langle\sigma_{\rm
[Fe/H]}\rangle = 0.23$ dex).

\subsection{The Reddenings\label{SECTION:reddening}}

	The reddening in the direction of each GC was determined using
the $E_{B\!-\!V}$ reddening maps of
Burstein~\&~Heiles\markcite{BH82}~(1982).  The uncertainties in the
$E_{B\!-\!V}$ values are approximately 0.01 mag due to uncertainties
in the reddening maps and the interpolation process.  The
$E_{B\!-\!V}$ values were converted to $E_{V\!-\!I}$ values using
$E_{V\!-\!I} = 1.36E_{B\!-\!V}$ (Taylor\markcite{T86}~1986,
Fahlman~\etal\markcite{F89}~1989).  G58 and G108 are located near NGC
205, which may lead to less accurate reddening estimates for these two
GCs.  An interstellar extinction of $A_V = (3.09\pm0.03)E_{B\!-\!V}$
(Rieke~\&~Lebofsky\markcite{RL85}~1985), where the quoted uncertainty
is the $2\sigma$ uncertainty, was adopted.  The comparison isochrone
for each GC was reddened by the appropriate amount for each GC and the
appropriate extinction corrections were applied.

	A summary of the properties of each GC is given in
Table~\ref{TABLE:gc_list}.  The GC names are from
Sargent~\etal\markcite{SK77}~(1977) except for Bo468, which is from
Battistini~\etal\markcite{BB87}~(1987).  The integrated $V$-band
magnitudes ($V$) and projected distances from the centre of M31 ($R$)
are from Huchra~\etal\markcite{HB91}~(1991).  The exception is the
coordinates and projected distance for Bo468, which are from
Battistini~\etal\markcite{BB87}~(1987).  The reference (column 8)
gives the source of the CMD for that GC\@.


\section{The Results\label{SECTION:results}}

	The distance modulus for each GC was found by computing the
weighted mean difference between the observed $V$-band magnitude of
the $i^{\rm th}$ fiducial point, $V_i$, and the absolute $V$-band
magnitude of the comparison isochrone (after correcting for reddening
and extinction) at the same $V\!-\!I$ colour as the fiducial point,
$M_V\Big({(V\!-\!I)}_i\Big)$.  The procedure for G280 and G351, for
which $B$- and $V$-band photometry is available, is the same except
that the $B\!-\!V$ colour index was used instead of the $V\!-\!I$ colour
index.  The formula for the distance modulus, $\mu_0$, and its formal
uncertainty, $\sigma_{\mu_0}$ is:

\begin{equation}
\mu_0 = \left(\sum_{i=1}^N {V_i \over w_i}
          - \sum_{i=1}^N {M_V\Big({(V\!-\!I)}_i\Big) \over w_i} \right)
        \Big/ \sum_{i=1}^N {1 \over w_i}
        \pm
        \sqrt{\sum_{i=1}^N {1 \over w_i}},
\label{EQUATION:mean_mu0}
\end{equation}

\noindent
where $w_i$ is the weight assigned to the $i^{\rm th}$ fiducial point
and $N$ is the number of fiducial points.  The weights were set to
$w_i = \sigma_{{V\!-\!I},i}^2 + \sigma_{V,i}^2$ where
$\sigma_{{V\!-\!I},i}$ and $\sigma_{V,i}$ are the uncertainties in the
location of the $i^{\rm th}$ fiducial point (see
\sect~\ref{SECTION:rgb}).  The formal uncertainties in the individual
distance moduli do not include uncertainties in the reddening or
uncertainties in the iron abundance estimates since these
uncertainties would be constant for a given GC\@.

\begin{figure}
\plotone{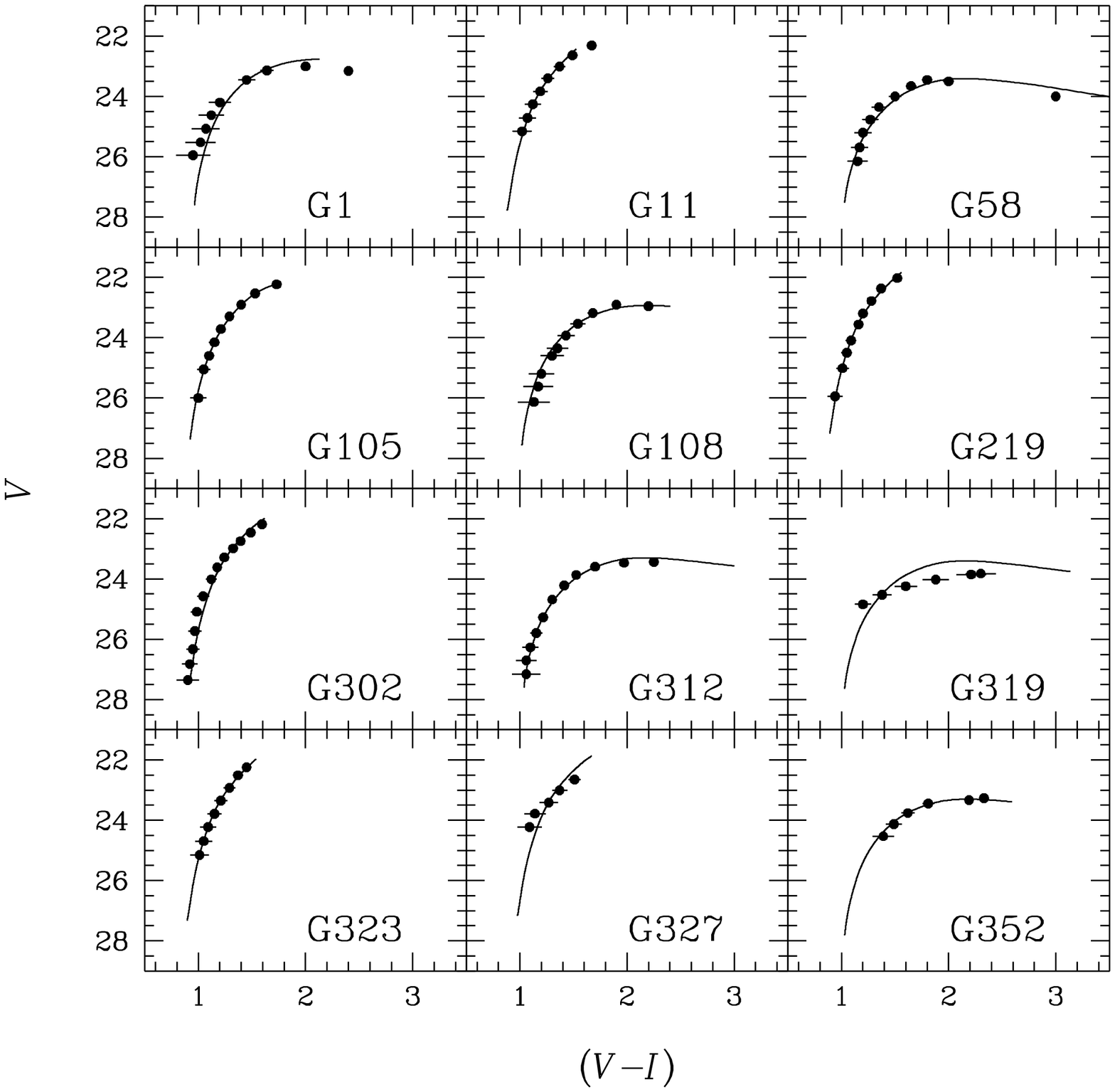}
\caption[Holland.f1.eps]{The RGB fiducial sequences for each GC with
$V$- and $I$-band data available. The best-fitting isochrones has been
overlaid on each RGB\@.  The isochrones have been reddened and
adjusted for interstellar extinction as described in
\sect~\ref{SECTION:reddening}.
\label{FIGURE:fits1}}
\end{figure}

\begin{figure}
\plotone{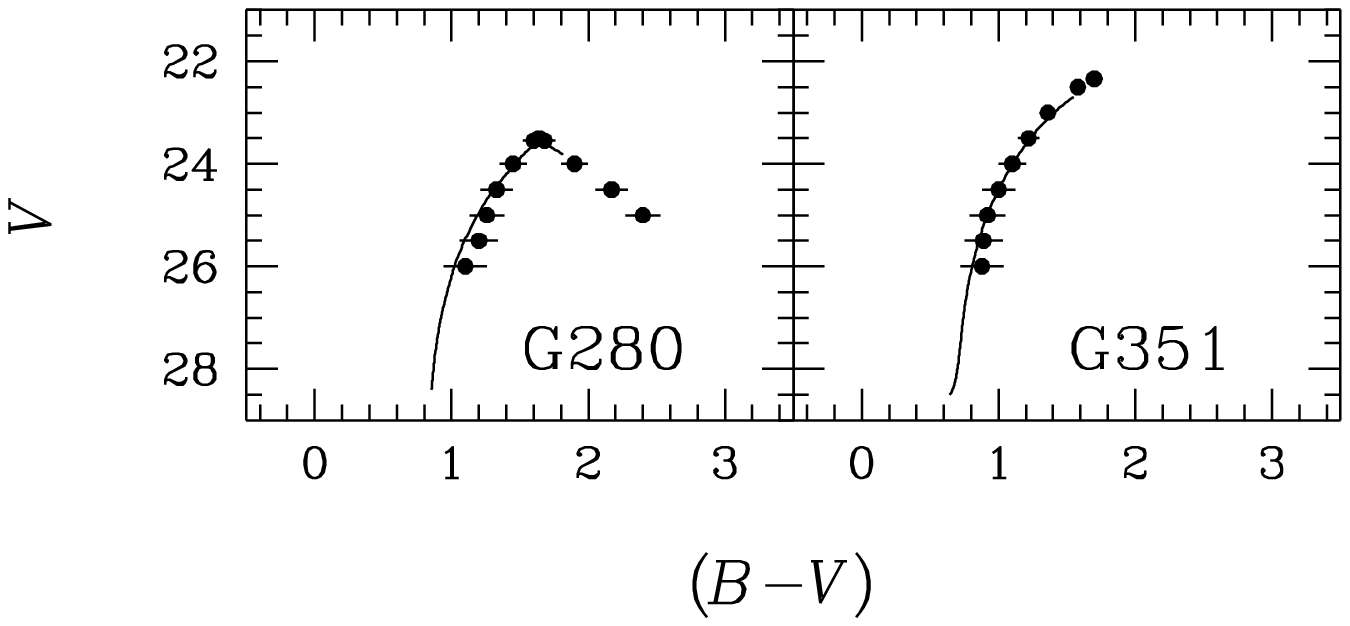}
\caption[Holland.f1.eps]{The RGB fiducial sequences for each GC with
$B$- and $V$-band data available. The best-fitting isochrones has been
overlaid on each RGB\@.  The isochrones have been reddened and
adjusted for interstellar extinction as described in
\sect~\ref{SECTION:reddening}.
\label{FIGURE:fits2}}
\end{figure}

	The uncertainty in $\mu_0$ for each GC due to the uncertainty
in the adopted reddening was estimated by recomputing $\mu_0$ using
$E_{B\!-\!V} + \sigma_{E_{B\!-\!V}}$ and $E_{B\!-\!V} -
\sigma_{E_{B\!-\!V}}$ then taking half of the difference between the
resulting $\mu_0$s.  The uncertainty in $\mu_0$ for each GC due to the
uncertainty in the spectroscopic iron abundance was estimated in the
same manner using $\FeH + \sigma_{\rm [Fe/H]}$ and $\FeH - \sigma_{\rm
[Fe/H]}$.  The results are presented in Table~\ref{TABLE:fits} along
with the quadratic sum of all three sources of error.
Table~\ref{TABLE:fits} also lists the best-fit values of the distance
modulus for each GC along with the uncertainties in each fit, the
root-mean-square (RMS) scatter in the fit, and the number of fiducial
points used to make the fit.
Figures~\ref{FIGURE:fits1}~and~\ref{FIGURE:fits2} shows the best-fits
between the fiducial sequences and the comparison isochrones.

	The uncertainties in the individual distance moduli are
dominated by the uncertainties in either the reddening estimate or
iron abundance determination.  An examination of the reddening maps of
Burstein~\&~Heiles\markcite{BH82}~(1982) shows that the spatial scale
of variations in reddening in the direction of M31 is larger than the
typical projected distances between the GCs.  This means that
uncertainties in the reddenings will not be completely independent
from one GC to another.  Uncertainties in the spectroscopic iron
abundances due to errors subtracting the continuum will result in
systematic errors in {\FeH}.  Recent results have suggested that the
Huchra~\etal\markcite{HB91}~(1991) iron abundances may be
systematically underestimated for the metal-rich GCs (Seitzer, private
communication; de~Freitas~Pacheco\markcite{dFP97}~1997;
Ponder~\etal\markcite{PB94}~1994).  Increasing the iron abundance in
GC stars will result in the upper RGB turning over to become flatter
and redder than the upper RGB of metal-poor GCs are.  Therefore,
underestimating the iron abundance of a GC will result in fitting a
comparison isochrone that is too bright, which will lead to the fitted
distance of that GC being larger than the true distance.  Therefore, a
systematic error in the measured iron abundance of a GC will lead to a
systematic error in the derived distance modulus of that GC that
increases as the iron abundance increases.

	Since the uncertainties have a systematic component, and the
size of this systematic component is unknown, they could not be used
to weight the individual distance moduli when computing the mean
distance modulus.  Therefore, unweighted distance moduli were used to
compute the distance to the centroid of the M31 GC system.  The
unweighted mean distance modulus for all the GCs in
Table~\ref{TABLE:fits} is $\langle\mu_0\rangle = 24.47 \pm 0.07$ mag.
All quoted uncertainties are standard errors in the means unless
otherwise stated.

	In order to test for effects arising from a possible
systematic under-estimate of the iron abundances of the iron-rich GCs
the sample of GCs was divided into two groups containing seven GCs
each.  The iron-poor group consisted of those GCs with $\FeH < -1.2$
dex while the iron-rich group was made up of those GCs with $\FeH >
-1.2$ dex.  The unweighted mean distance moduli computed for each
group were $\langle\mu_0\rangle = 24.41 \pm 0.14$, for $\FeH < -1.2$,
and $\langle\mu_0\rangle = 24.53 \pm 0.06$, for $\FeH > -1.2$.  The
Student's $t$-test gives a probability of obtaining these two means by
chance from a single distribution of 42.87\%.  Therefore, there is no
evidence that the derived distances of the iron-rich GCs are greater
than those of the iron-poor GCs.

	The quality of the individual CMDs that were used to obtain
the RGB fiducial sequences varies greatly.  In addition, the
photometry was obtained using three different instruments and
calibrated using different procedures (see CR95, FP96, and HFR for
details).  In order to eliminated any systematic effects arising from
the combination of several different data sets the unweighted mean of
the distance moduli of the GCs observed with the WFPC2 and calibrated
by FP96 was computed.  This is the largest self-consistent subset of
GCs ($N = 5$) available.  The unweighted mean is $\langle\mu_0\rangle
= 24.37 \pm 0.06$, which is within 1.5$\sigma$ of the mean of the
entire sample of GCs.


\section{Conclusions\label{SECTION:conc}}

	The distances to fourteen GCs in the M31 system have been
estimated by comparing the observed RGB of each GC to a theoretical
isochrone with the same iron abundance.  The unweighted mean distance
modulus of the GCs is $\langle\mu_0\rangle = 24.47 \pm 0.07$ mag.  If
only the ``best'' data set (the largest data set that used the same
instrument and calibrations) is used a distance modulus of
$\langle\mu_0\rangle = 24.37 \pm 0.06$ mag is obtained.  These
estimates of the distance to M31 are consistent with distances
obtained using other techniques and do not support the need for a
revision of the distance to M31.

	This method of determining the distance to M31 depends on
accurate iron abundance estimates for the individual GCs and needs a
sample of M31 GCs with high-precision photometry of their RGB stars.
In order to improve the accuracy of this distance estimator deep
($V_{\lim} \gtrsim 26$) photometry will be needed for a much larger
sample of M31 GCs.  In addition, further work needs to be done to
reduce the uncertainties in the spectroscopic iron abundance
measurements of M31 GCs.


\acknowledgments

	This research is based, in part, on observations made with the
NASA/ESA {\sl Hubble Space Telescope\/} obtained at the Space
Telescope Science Institute.  STScI is operated by the Association of
Universities for Research in Astronomy Inc.\ under NASA contract
NAS5--26555.  This research is also based, in part, on observations
made with the Canada--France--Hawaii Telescope operated by the
National Research Council of Canada, the Centre National de la
Recherche Scientifique de France and the University of Hawaii.  The
author would like to thank Don VandenBerg for kindly providing copies
of the isochrones used in this study.



\newpage


%
%


\begin{deluxetable}{lcccrccc}
\tablewidth{0 pt}
\tablenum{1}
\tablecaption{Properties of the globular clusters.\label{TABLE:gc_list}}
\tablehead
{\colhead{Cluster} &
 \colhead{$\alpha_{\rm J2000}$} &
 \colhead{$\delta_{\rm J2000}$} &
 \colhead{$V$} &
 \colhead{$R$} &
 \colhead{[Fe/H]} &
 \colhead{$E_{B\!-\!V}$} &
 \colhead{Ref.}
}
\startdata
G1    & $00^{\rm h} 32^{\rm m} 46\fs8$ & $+39\arcdeg 34\arcmin 42\arcsec$ & 13.70 & $152\farcm3$ & $-1.08\pm0.09$ & 0.06 & 2 \nl
G11   & $00^{\rm h} 36^{\rm m} 20\fs5$ & $+40\arcdeg 53\arcmin 38\arcsec$ & 16.39 &  $75\farcm7$ & $-1.89\pm0.17$ & 0.06 & 1 \nl
G58   & $00^{\rm h} 40^{\rm m} 26\fs8$ & $+41\arcdeg 27\arcmin 28\arcsec$ & 15.80 &  $28\farcm2$ & $-0.57\pm0.15$ & 0.06 & 2 \nl
G105  & $00^{\rm h} 41^{\rm m} 42\fs2$ & $+40\arcdeg 12\arcmin 23\arcsec$ & 16.35 &  $64\farcm8$ & $-1.49\pm0.17$ & 0.06 & 2 \nl
G108  & $00^{\rm h} 41^{\rm m} 43\fs3$ & $+41\arcdeg 33\arcmin 32\arcsec$ & 15.80 &  $20\farcm8$ & $-0.94\pm0.27$ & 0.09 & 2 \nl
Bo468 & $00^{\rm h} 43^{\rm m} 12\fs5$ & $+39\arcdeg 47\arcmin 57\arcsec$ & 18.12 &  $87\farcm0$ & \nodata        & 0.06 & 2 \nl
G219  & $00^{\rm h} 43^{\rm m} 18\fs0$ & $+39\arcdeg 49\arcmin 14\arcsec$ & 15.10 &  $87\farcm2$ & $-1.83\pm0.22$ & 0.06 & 2 \nl
G280  & $00^{\rm h} 44^{\rm m} 29\fs9$ & $+41\arcdeg 21\arcmin 37\arcsec$ & 14.30 &  $20\farcm5$ & $-0.70\pm0.12$ & 0.06 & 2 \nl
G302  & $00^{\rm h} 45^{\rm m} 25\fs2$ & $+41\arcdeg 05\arcmin 30\arcsec$ & 14.90 &  $32\farcm1$ & $-1.76\pm0.18$ & 0.08 & 3 \nl
G312  & $00^{\rm h} 45^{\rm m} 58\fs8$ & $+40\arcdeg 42\arcmin 32\arcsec$ & 16.05 &  $48\farcm9$ & $-0.70\pm0.35$ & 0.08 & 3 \nl
G319  & $00^{\rm h} 46^{\rm m} 22\fs1$ & $+40\arcdeg 17\arcmin 00\arcsec$ & 15.75 &  $72\farcm1$ & $-0.66\pm0.22$ & 0.07 & 1 \nl
G323  & $00^{\rm h} 46^{\rm m} 33\fs6$ & $+40\arcdeg 44\arcmin 14\arcsec$ & 17.09 &  $53\farcm8$ & $-1.96\pm0.29$ & 0.07 & 1 \nl
G327  & $00^{\rm h} 46^{\rm m} 49\fs4$ & $+42\arcdeg 44\arcmin 49\arcsec$ & 16.00 &  $99\farcm7$ & $-1.76\pm0.11$ & 0.12 & 1 \nl
G351  & $00^{\rm h} 49^{\rm m} 33\fs1$ & $+41\arcdeg 35\arcmin 32\arcsec$ & 15.18 &  $79\farcm0$ & $-1.80\pm0.31$ & 0.09 & 2 \nl
G352  & $00^{\rm h} 50^{\rm m} 10\fs0$ & $+41\arcdeg 41\arcmin 01\arcsec$ & 16.01 &  $87\farcm1$ & $-0.85\pm0.33$ & 0.09 & 1 \nl
\enddata
\tablerefs{(1)~Couture~et~al.\markcite{CR95}~(1995);
(2)~Fusi~Pecci~et~al.\markcite{FP96}~(1996);
(3)~Holland~et~al.\markcite{HF97}~(1997).
}
\end{deluxetable}




%
%


\begin{deluxetable}{lccccccr}
\tablewidth{0 pt}
\tablenum{2}
\tablecaption{The best-fitting distance moduli for each GC.\label{TABLE:fits}}
\tablehead
{\colhead{Cluster} &
 \colhead{$\mu_0$} &
 \colhead{$\sigma_{\mu_0}$(fit)\tablenotemark{a}} &
 \colhead{$\sigma_{\mu_0}$($E_{B\!-\!V}$)\tablenotemark{b}} &
 \colhead{$\sigma_{\mu_0}$([Fe/H])\tablenotemark{c}} &
 \colhead{$\sigma_{\mu_0}$(total)\tablenotemark{d}} &
 \colhead{RMS} &
 \colhead{$N$}
}
\startdata
G1    & 24.53 & 0.03 & 0.05 & 0.07 & 0.09 &  0.448 &  8 \nl
G11   & 24.80 & 0.03 & 0.11 & 0.07 & 0.13 &  0.078 &  7 \nl
G58   & 24.39 & 0.02 & 0.06 & 0.06 & 0.09 &  0.204 & 10 \nl
G105  & 24.35 & 0.01 & 0.09 & 0.31 & 0.32 &  0.089 &  9 \nl
G108  & 24.40 & 0.02 & 0.04 & 0.57 & 0.57 &  0.236 & 10 \nl
G219  & 24.18 & 0.01 & 0.11 & 0.02 & 0.11 &  0.071 &  9 \nl
G280  & 24.86 & 0.04 & 0.05 & 0.09 & 0.11 &  0.300 &  8 \nl
G302  & 24.23 & 0.01 & 0.10 & 0.03 & 0.10 &  0.386 & 11 \nl
G312  & 24.42 & 0.01 & 0.07 & 0.21 & 0.22 &  0.140 & 11 \nl
G319  & 24.49 & 0.05 & 0.08 & 0.21 & 0.23 &  0.429 &  6 \nl
G323  & 24.31 & 0.03 & 0.11 & 0.04 & 0.12 &  0.037 &  8 \nl
G327  & 23.98 & 0.04 & 0.11 & 0.13 & 0.17 &  0.505 &  5 \nl
G351  & 25.01 & 0.04 & 0.08 & 0.15 & 0.17 &  0.293 &  7 \nl
G352  & 24.64 & 0.03 & 0.04 & 0.18 & 0.19 &  0.070 &  6 \nl
\enddata
\tablenotetext{a}{The formal uncertainty in the fit, from
Equation~1.}
\tablenotetext{b}{The uncertainty due to uncertainties in the reddening.}
\tablenotetext{c}{The uncertainty due to uncertainties in the iron abundance.}
\tablenotetext{d}{The total uncertainty in $\mu_0$.}
\end{deluxetable}




\newpage

\end{document}